# High-pressure growth effect on the properties of high-$T_c$ iron-based superconductors: A short review

Priya Singh[1], Manasa Manasa[1], Mohammad Azam[1], Shiv J. Singh[1*]

[1]*Institute of High Pressure Physics (IHPP), Polish Academy of Sciences, Sokołowska 29/37, 01-142 Warsaw, Poland*

*Corresponding author:

Email: sjs@unipress.waw.pl

https://orcid.org/0000-0001-5769-1787

# Abstract

The high-pressure growth technique is a vital approach that facilitates the stabilization of new phases and allows for meticulous control of structural parameters, which significantly impact electronic and magnetic properties. We present a short review of our ongoing investigations into various families of iron-based superconductors (IBS), employing the high-gas pressure and high-temperature synthesis (HP-HTS) method. This technique is capable of producing the gas pressures up to 1.8 GPa and a heating temperature of up to 1700 °C through a three-zone furnace within a cylindrical chamber. Different kinds of IBS samples are prepared using HP-HTS and characterized through various measurements to reach the final conclusions. The results demonstrate that the high-pressure growth technique significantly enhances the properties of IBS, including the transition temperature, critical current density, and pinning force. In addition, the quality of the samples and their density are improved through the intergrain connections. Furthermore, the comprehensive evaluations and investigations prove that a growth pressure of 0.5 GPa is sufficient for producing high-quality IBS bulks under the optimized synthesis conditions.

# 1. Introduction

The discovery of iron-based superconductors occurred in 2008 [1], leading to the identification of over 100 compounds within this high-$T_c$ family. According to the crystal structure of the parent compounds, these high-$T_c$ materials can be classified into six distinct families [2], [3], [4], [5]: $RE$FeAsO ($RE$1111; $RE$ = rare earth), $A$Fe$_2$As$_2$ (122; $A$ = Ba, K, Ca), (Li/Na)FeAs (111), thick perovskite-type oxide blocking layers, such as Sr$_4$V$_2$O$_6$Fe$_2$As$_2$ (42622), Sr$_4$Sc$_2$O$_6$Fe$_2$P$_2$ (42622) and chalcogenide Fe$X$ representing 11 ($X$ = chalcogenide). The parent compound of the 1111 family exhibits a structural and magnetic transition around 150 K, and appropriate doping enables the observation of superconductivity [6]. The superconducting transition ($T_c$) of 1111 was increased and reached a maximum value of 58 K for F-doped Sm1111, which is the highest value for IBS [7], [8]. On the other hand, there are some stoichiometric families associated with IBS, including the 111 and 1144 families. The stoichiometric family 1144 does not exhibit the structural and magnetic transition; however, the highest $T_c$ of 34-35 K is achieved without doping [9], [10] for IBS. Furthermore, the 11 family represents the simplest family of IBS, providing the $T_c$ of 8 K [11]. However, with 50% Te doping at Se sites, the $T_c$ can be increased up to 15 K [12]. One of the basic problems of this high $T_c$ superconductor is the preparation of the samples in a pure phase with well-connected grain boundaries. For example, 1111 family consistently exhibits the impurity phases ($RE$O and SmAs) during the growth process by the conventional synthesis process at the ambient pressure (CSP-AP), and maximum doping can be possibly restricted to the optimal region, i.e., ~20% [13]. Exceeding 20% doping appears mainly in the form of the impurity phases, leading to rapid degradation of the superconducting properties [6]. In the case of the 11 family, the tetragonal 11 (FeSe) phase exhibits the superconductivity and it is crucial to maintain this pure tetragonal phase during the growth process, avoiding any impurity phases such as the hexagonal phase of 11 [14, 15]. Hence, preparing a completely clean tetragonal 11 phase through a convenient synthesis process at ambient pressure is a challenging task. Similar issues are also observed with other families of this high-$T_c$ IBS [6].

Many studies have been reported regarding the effects of applied external pressure on IBS, where the superconducting properties can be enhanced [16]. For example, F-doped LaFeAsO (La1111) prepared by CSP-AP demonstrates a $T_c$ ~ 26 K at ambient pressure [1], which can be increased up to 43 K at 3 GPa [17]. On the other hand, very few studies have been performed on the high-pressure growth process, demonstrating its potential effectiveness [18, 19]. The development of the 1111 single crystal cannot be achieved through CSP-AP. For the first time, Karpinski et al. employed the high-pressure growth technique on various doping contents of the 1111 family and reported high-quality single crystals [18], [20] that exhibit the superconducting properties. Nonetheless, these crystals were tiny and were not appropriate for transport and angle-resolved photoemission spectroscopy (ARPES) measurements. Weiss et al. [19] conducted a high-pressure synthesis of bulk $(Ba_{0.6}K_{0.4})Fe_2As_2$ (BaK122) and $Ba(Fe_{0.92}Co_{0.08})_2As_2$ (BaCo122), using high-impact ball milling. These bulks have the higher phase purity and the highest bulk $J_c$ (0.1 MA·cm$^{-2}$, 0 T) value with the enhanced the irreversibility field ($H_{irr}$) in comparison to other bulk BaK122 or BaCo122 samples prepared by CSP-AP. The advancement of polycrystalline powder synthesis is crucial for enhancing $J_c$, and the processes involved in wire fabrication have also influenced their performance. Moreover, the investigations involving 122 families have demonstrated that the enhanced sample densification through high-pressure sintering, along with the texturing of grains in the wire core during the drawing process, significantly improves $J_c$ [21], [22]. These investigations inspire us to concentrate on this research area and enhance the quality and size of the IBS samples.

The last 16 years of discovery of IBS suggest that the CSP-AP approach is inadequate for addressing the sample difficulties of IBS [6]. High-quality samples are essential to comprehend the intrinsic features of these high-temperature superconductors. Consequently, we are currently concentrating on the high-pressure growth of IBS utilizing the high gas pressure and high-temperature synthesis (HP-HTS) technique [23] to optimize the growth parameters for various families of IBS, particularly the 1111 [24], [25], [26], 1144 [27] [28], and 11 [15], [14], [29] families. These families are selected to investigate the impact of high-pressure growth on doped and stoichiometric IBS families. In this short review, we have summarized our main finding from high-pressure growth effects on these IBS families. We anticipate that this study will significantly aid the research community in addressing the fundamental issues associated with these remarkable high-temperature superconductors.

## 2. High-Pressure Technique

We have used the HP-HTS technique for the synthesis of IBS, and the block diagram of this technique is illustrated in Figure 1 [23]. This approach is based on the hot isostatic pressing (HIP) technique [30], [31], through which we can achieve an inert gas pressure of 1.8 GPa and a temperature of 1700 °C with precise temperature and pressure control in the sample space. This technique employs three oil-based pistons arranged in series to achieve the required high pressure, adhering to the principle of Boyle's Law [32], as depicted in Figure 1. The first piston has a gas pressure of 0.02 GPa from the gas bottle, and as it moves upward, a gas pressure up to 0.08 GPa is created for the second piston chamber. Subsequently, the second piston moves in an upward direction, transferring the gas pressure up to 0.4 GPa for the third piston chamber and high-pressure sample chamber. The third piston plays a crucial role in reaching the ultimate pressure, which is a little bigger in size and creates a pressure of up to 1.8 GPa. Furthermore, the HP-HTS system offers a large sample space of up to 15 cm$^3$, providing the feasibility to grow large crystals and a large amount of bulks in one batch [23].

## 3. Effects on the properties of IBS:

### *3.1 11 family:*

The simplest iron-based superconductors are FeSe, which belong to the 11 family and exhibit the superconductivity at ~8 K. Through an appropriate doping i.e., chemical pressure, such as Te doping at Se sites, the $T_c$ can reach up to 15 K [12], [33], [34] while by the applied external pressure, $T_c$ can increase up to 36-37 K. The crystal structure of this family is the simplest in all of IBS, but the synthesis phase diagram of FeSe is very complicated because it has many different stable phases [35], such as hexagonal $Fe_7Se_8$, monoclinic $Fe_3Se_4$, orthorhombic $FeSe_2$, hexagonal $\delta$-$Fe_xSe$ and tetragonal $\beta$-$Fe_xSe$ phase [36]. Interestingly, in all these phases, only the tetragonal structure of FeSe exhibits the superconductivity with a $T_c$ of 8 K at ambient pressure [11], [37]. The presence of these stable phases consistently complicates the preparation of a pure superconducting phase during the growth of single crystals [38], [39] or polycrystalline samples [12], [33], [40], [41, 42]. Certain stable phases, notably hexagonal $\delta$-$Fe_xSe$ and hexagonal $Fe_7Se_8$, often coexist with the main tetragonal $\beta$-$Fe_xSe$ phase during the synthesis process, but they are not favourable for the superconducting properties [43], [44], [45], [14], [46]. Recent studies have suggested that conventional synthesis processes at ambient pressure are ineffective in enhancing the key critical current characteristics, grain connectivity, and phase purity of bulk samples for practical applications. Different synthesis routes have been reported for the 11 family and to produce Fe(Se,Te) samples at elevated temperatures ranging from 880 to 1000 °C for extended durations [44], [47], [12]. Despite numerous annealing procedures to alter the sample properties, the process was still unable to reduce the foreign phases during the growth of high-quality samples of 11 family. To overcome this problem, we have utilized the HP-HTS technique to optimize the growth process of 11 family by preparing various Fe(Se,Te) bulks under different synthesis conditions. Table 1 lists the sample codes and the synthesis conditions for the few samples prepared using HP-HTS and CSP-AP processes. Additional information regarding these experiments and other samples is provided in our previously published works [15], [29]. We have selected the optimal doping content for the 11 family to optimize the synthesis conditions by HP-HTS process, specifically 50% Te doping at Se sites, i.e., FeSe$_{0.5}$Te$_{0.5}$, which provides the highest $T_c$ of 14-15 K.

Since there were no reports on the high-pressure growth process of Fe(Se, Te), it was essential to optimize this process. We have synthesized FeSe$_{0.5}$Te$_{0.5}$ bulks under various conditions, including growth pressure and durations, as well as utilizing the samples sealed into a Ta-tube or without a Ta-tube. These samples are designated as HIP-S1 to HIP-S13, corresponding to various growth pressures ranging from 0 to 1 GPa, as illustrated in Figure 2, and more details are reported elsewhere [15]. All the prepared samples are characterized by structural and microstructural analysis, transport, and magnetic measurements [15]. The obtained results are plotted for various FeSe$_{0.5}$Te$_{0.5}$ bulks prepared under different growth pressures in Figure 2, and the details about some samples are mentioned in Table 1. The hexagonal phase decreased to its lowest level for certain prepared samples at 0.5 GPa for 1 h, using both ex-situ and in-situ methods, as illustrated in Figure 2(a), thereby promoting the formation of the tetragonal phase of FeSe$_{0.5}$Te$_{0.5}$. The $T_c$ has varied, as shown in Figure 2(b) for the different synthesis pressures. The samples HIP-S9 prepared by ex-situ process (with sealing into a Ta-Tube) at 0.5 GPa and HIP-S11 prepared by *in-situ* process (sealed into Ta-Tube) at 0.5 GPa demonstrate an increase in $T_c$ of up to 17.2 K accompanied by a reduced hexagonal phase, akin to that of the parent compound. The $T_c$ is enhanced by 2-3 K compared to FeSe$_{0.5}$Te$_{0.5}$ prepared by the CSP method. Figure 2(c) shows the critical current density $J_c$ at 7 K for magnetic fields 0 T and 5 T for the three samples and the parent compound. Interestingly, the $J_c$ of HIP-S9 is enhanced by one order of magnitude compared to the parent sample. We also noticed that the $J_c$ value increased at higher magnetic fields, which may be due to the high-pressure growth procedure strengthening the pinning and grain connections. This is confirmed by the analysis of pinning force at 5 T for the sample HIP-S9, as shown in Figure 2(d), suggesting a strong pinning nature and an improvement in the $J_c$ [15].

Basically, during the high-pressure growth of the 11-phase, the tetragonal and hexagonal phases compete with each other. Therefore, it is necessary to identify the appropriate synthesis conditions to minimize the hexagonal phase. When the hexagonal phase reduces, the formation of the tetragonal phase increases. The hexagonal phase reduces to the minimum level at 0.5 GPa under the suitable synthesis conditions (Figure 2(a)), which worked well for both *ex-situ* and *in-situ* processes [15], [29]. On the other hand, FeSe$_{0.5}$Te$_{0.5}$ bulks prepared at different synthesis pressures (i.e., 0.3, 0.7, and 1 GPa) have low performance with the superconducting properties due to the formation of the hexagonal phase (Figure 2). Interestingly, even using a high synthesis pressure (>0.5 GPa) and heating for a long duration is not able to reduce the hexagonal phase during the preparation of Fe(Se,Te) bulks. The detailed analysis of high-pressure growth effects on 11 family reveals that the optimal pressure (0.5 GPa for 1 h) can reduce the hexagonal phase to a minimum level and promote the formation of the tetragonal phase. While other synthesis pressure somehow promotes the hexagonal phase by reducing the superconducting tetragonal phase. To gain a better understanding of the intriguing behaviour of the hexagonal and tetragonal phases of 11, a more in-depth examination of the phase formation of Fe(Se,Te) under high-pressure growth effects is essential. In summary, our results [15], [29] reveal that a growth pressure of 0.5 GPa is adequate to achieve a nearly pure tetragonal phase of $FeSe_{0.5}Te_{0.5}$, which exhibits the enhanced superconducting properties.

## *3.2 1144 family:*

The stoichiometric compound $CaKFe_4As_4$ belonging to the 1144 family, which was discovered in 2016, offers a maximum $T_c$ of 34-35 K without doping [9, 48, 10]. Our aim was to investigate the effects of the high-pressure growth on this stoichiometric family. The requisite to observe the high superconducting properties is the phase purity and well-connected grain boundaries [9], [10], which is highly challenging to achieve by CSP-AP [10]. The synthesis of $CaKFe_4As_4$ has been reported utilizing a spark plasma sintering (SPS) technique [49], resulting a $T_c^{\text{onset}}$ of 35 K and a calculated $J_c$ of $8.1 \times 10^4$ $A/cm^2$ at 5 K and 0 T [49]. The density of this SPS sample was approximately 100%, but the detection of several impurity phases led to a significantly lower $J_c$ compared to the previously reported 1144 single crystals [48]. To address this issue, our group reported the high-pressure synthesis of $CaKFe_4As_4$ for the first time by considering various growth parameters [28], [27].

Our previous research involving 11 families indicated that a synthesis pressure of 0.5 GPa for 1 h is sufficient for producing the high-quality bulk material [15]. Consequently, we employed this optimal pressure and duration (0.5 GPa and 1 h) to synthesize several $CaKFe_4As_4$ samples at 0.5 GPa under diverse conditions as listed in Table 1. Since the initial precursors, such as K and As, are very reactive, we had decided to use the ex-situ process, i.e., the prepared 1144 bulks by CSP-AP were used for HP-HTS process at 0.5 GPa for 1 h either into an open or closed Ta-tube. Here, the two samples prepared by HP-HTS are discussed with the parent (Table 1): The first sample, i.e., HIP_1, involves the parent 1144 bulks within an open Ta-tube. The second sample, i.e., HIP_2, where the parent 1144 was placed into the pressure chamber into an open Ta-tube, as first step and then a sealed Ta-tube in the second step. A more detailed study is reported in our published papers [28], [27], and the main results of our investigation are illustrated in Figure 3. Figure 3(a) suggests the enhancement of $T_c$ by 2 K for the sample HIP_1 prepared by HP-HTS, where the growth conditions of HIP_2 samples are not suitable for the superconducting properties of 1144, even though the samples are prepared at 0.5 GPa. Moreover, the $J_c$ of HIP_1 has also been enhanced by one order of magnitude with and without the magnetic fields, as shown in Figure 3(b). We have calculated the pinning force for these samples in Figure 3(c), and found that it increases for the HIP_1 compared to the parent and HIP_2, suggesting the strong pinning characteristics for the sample HIP_1. As a result, the enhancement of $J_c$ is observed across the entire magnetic field range for HIP_1. This $J_c$ and pinning force enhancement could be due to the improvement of material density, grain connections, and the appropriate pinning centers as reported elsewhere [50], [16]. It suggests that high-pressure synthesis performs effectively to improve the intergrain connections and the pinning properties of 1144 superconductors. Actually, we prepared the HIP_2 sample in two steps: first, we placed it into an open Ta-tube, and then we sealed it into a Ta-tube (Table 1). Each step involved placing the sample in a high-pressure chamber and preparing it under the optimal conditions (0.5 GPa for 1 h). We have observed a little amount of the impurity phases (especially, $CaFe_2As_2$) for HIP_2 compared to HIP_1 and parent samples [27]. Since $CaKFe_4As_4$ (1144) phase is very sensitive to the stoichiometric composition, even a small change in stoichiometry can cause the impurity phases ($CaFe_2As_2$ and $KFe_2As_2$), for example due to the evaporation of the lighter elements, especially potassium (K). We believe that preparing 1144 under high pressure into an open Ta-tube (one step only), i.e., HIP_1 sample (Table 1), prevents the evaporation of lighter elements like potassium (K) under the applied pressure and results in the formation of a pure 1144 phase. However, when this 1144 sample was prepared into an open and then a closed Ta-tube through a two-step process (like the sample HIP_2) inside a high-pressure chamber of HP-HTS, there could be a small change in the stoichiometry because of a small amount of K evaporation. This could be possible due to the long heating time (total: 2 h, including two-step process). As a result, it produces a small amount of $CaFe_2As_2$ as an impurity phase for HIP_2. Due to this possible reason and on the basis of the observed superconducting properties, we have concluded that the growth pressure of 0.5 GPa is inappropriate for the preparation conditions of HIP_2 sample. It is also well observed from the study of $FeSe_{0.5}Te_{0.5}$ bulk [15], [29] that long heating time under high pressure also reduces the sample's quality and superconducting properties. Furthermore, our study of the 1144 family confirms that the applied growth pressure of 0.5 GPa works as an optimal growth pressure, which supports our finding for 11 family. All of these results suggest that $CaKFe_4As_4$ prepared at 0.5 GPa by HP-HTS process into an open Ta-tube exhibits high superconducting properties compared to samples prepared under different conditions [27, 28]. Hence, the high-pressure synthesis has been promising for the 1144 family, and further studies are required in this direction for the improvement of the superconducting properties.

## *3.3 1111 family:*

The 1111 family associated with IBS exhibits the highest $T_c$ of 58 K through F-doped SmFeAsO (Sm1111) [7], [8]. However, it is challenging to prepare high-quality and large-size samples without the impurity phases for this interesting family. Various reports based on the 1111 family suggest that CSP-AP is inadequate to improve the sample's quality as well as the superconducting properties, necessitating the implementation of a novel methodology, such as the high-pressure technique. Very few studies have depicted the positive effects of the high-pressure technique on the growth of 1111 single crystals [51], [52], [53]. Furthermore, our recent studies based on the HP-HTS method have also been reported for Fe(Se,Te) (11) and $CaKFe_4As_4$ (1144) families, which improve the sample's quality as well as also enhance the superconducting properties [15], [29], [27]. These studies encouraged us to implement the HP-HTS technique for this intriguing 1111 family, especially for F-doped Sm1111.

We have optimized the synthesis conditions of F-doped Sm1111 ($SmFeAsO_{1-x}F_x$) by preparing bulks with an optimal fluorine doping content ($x = 0.2$) under various growth parameters using the HP-HTS at various applied pressures up to 1 GPa. These studies are compared with the parent sample ($SmFeAsO_{0.8}F_{0.2}$) prepared using CSP-AP to understand the effects of high-pressure growth on its superconducting properties [24]. All samples were thoroughly well characterized by structural, microstructural, Raman spectroscopy, transport, and magnetic measurements to reach the conclusion. We have prepared different batches of $SmFeAsO_{0.8}F_{0.2}$ bulks by considering various synthesis conditions and here, one specific batch is focused on, referred to as G-batch samples and depicted in Figure 4. More details about the samples are listed in Table 1 and also given elsewhere [24]. The sample labelled as G0 was prepared by sealing it into a Ta-tube for 1 h under the ambient pressure (0 GPa), while the other four samples, G1, G2, G3, and G4, were prepared into the sealed Ta-tubes at varying pressures of 0.3 GPa, 0.5 GPa, 0.7 GPa, and 1 GPa, respectively (Table 1). Notably, all the samples prepared by HP-HTS have no substantial improvement in $T_c$ and $J_c$ as compared to the parent sample P, as reported in the reference [24]. In Figures 4(a) and 4(b), a slight improvement in $J_c$ is observed for the optimal sample (G2) from the G-batch compared to the parent sample (P), but the $T_c$ is almost the same for both samples. These enhanced properties might be due to a small improvement (~8-9%) in the sample density compared to the parent P [24]. At the same time, the flux pinning behaviour for the sample G2 at 0.5 GPa is also improved, as shown in Figure 4(c), which stimulates the improvement of the $J_c$, similar to the 1144 and 11 families. Our studies suggest that the amount and type of impurity phases were almost the same for all samples (G0 to G4), similar to the parent compound (P) despite the application of the growth pressure up to 1 GPa; thus, the high synthesis pressure was not able to reduce the impurity phases. Due to this, the observed superconducting properties are almost the same for all samples (P, G0 to G4). Hence, the sample quality and superconducting properties of F-doped Sm1111 have moderate variations [24]. Further studies are required in this direction to enhance the sample's quality and superconducting properties of the 1111 family.

## 4. Conclusion

Studies on high-pressure growth utilizing various IBS families demonstrate that an optimal growth pressure of 0.5 GPa for 1 h enhances the superconducting properties in comparison to other growth conditions. High-pressure synthesis of $CaKFe_4As_4$ and Fe(Se,Te) from the 1144 and 11 families has enhanced the $J_c$ by one order of magnitude and raised the $T_c$ by 2-3 K. Nonetheless, SmFeAs(O, F) belonging to the 1111 family has demonstrated distinct outcomes in contrast to the 11 and 1144 families, where the sample's quality and nature of impurities were nearly identical across all samples produced by HP-HTS and CSP-AP methods. As a result, there was no significant improvement in the superconducting properties and sample quality of the 1111 family. Nevertheless, the HP-HTS effects exhibit a small enhancement in the $J_c$ and pinning behaviors for the optimal SmFeAs(O, F) samples. Our investigations prove that high-pressure growth processes are effective in improving the sample quality as well as the superconducting properties of these high-$T_c$ superconductors; nonetheless, more research is required in this field. This approach is expected to enhance the investigation of IBS materials, resulting in the improved sample quality, advancements in superconducting properties, and the development of practical applications like tapes and wires.

**Acknowledgments:**

The work was supported by SONATA-BIS 11 project (Registration number: 2021/42/E/ST5/00262) funded by National Science Centre (NCN), Poland. SJS acknowledges financial support from National Science Centre (NCN), Poland through research Project number: 2021/42/E/ST5/00262.

**Table 1:** A list of the sample codes and synthesis conditions for the 11, 1144, and 1111 families. More details about the synthesis conditions are mentioned in the references [15], [24], [27]. "Sealed into a Ta-tube" means the sample was sealed into a Ta-tube by ARC melter, and "Open Ta-tube" means the sample was placed into the Ta-tube without sealing, as a crucible.

| Sample code | Synthesis conditions |
|---|---|
| Parent ($FeSe_{0.5}Te_{0.5}$) | *First step*: 600 °C, 11 h, 0 MPa (without Ta-tube) ↓ *Second step*: 600 °C, 4 h, 0 MPa (without Ta-tube) |
| HIP-S9 ($FeSe_{0.5}Te_{0.5}$) | *First step*: 600 °C, 11 h, 0 MPa (without Ta-tube) ↓ *Second step*: 600 °C, 1 h, 500 MPa (sealed into a Ta-tube) |
| HIP-S11 ($FeSe_{0.5}Te_{0.5}$) | *First step*: 600 °C, 1 h, 500 MPa (sealed into a Ta-tube) |
| Parent ($CaKFe_4As_4$) | *First step:* heated at 955°C, 6 h, 0 MPa (sealed into a Ta-tube) ↓ *Second step:* heated at 955°C, 2 h, 0 MPa (sealed into a Ta-tube) |
| HIP_1 ($CaKFe_4As_4$) | *First Step* : 500 °C, 1 h, 500 MPa (Open Ta-tube) |
| HIP_2 ($CaKFe_4As_4$) | *First step*: 500 °C, 1 h, 500 MPa (Open Ta-tube) ↓ *Second step*: 500 °C, 1 h, 500 MPa (sealed into a Ta-tube) |
| P ($SmFeAsO_{0.8}F_{0.2}$) | *First step* : 900 °C, 45 h, 0 MPa (Open Ta-tube) |
| G0 ($SmFeAsO_{0.8}F_{0.2}$) | *First step*: 900 °C, 45 h, 0 MPa (Open Ta-tube) ↓ *Second step*: 900 °C, 1 h, 0 GPa (sealed into a Ta-tube) |
| G1 ($SmFeAsO_{0.8}F_{0.2}$) | *First step*: 900 °C, 45 h, 0 MPa (Open Ta-tube) ↓ *Second step*: 900 °C, 1 h, 0.3 GPa (sealed into a Ta-tube) |
| G2 ($SmFeAsO_{0.8}F_{0.2}$) | *First step*: 900 °C, 45 h, 0 MPa (Open Ta-tube) ↓ *Second step*: 900 °C, 1 h, 0.5 GPa (sealed into a Ta-tube) |
| G3 ($SmFeAsO_{0.8}F_{0.2}$) | *First step*: 900 °C, 45h, 0 MPa (Open Ta-tube) ↓ *Second step*: 900 °C, 1 h, 0.7 GPa (sealed into a Ta-tube) |
| G4 ($SmFeAsO_{0.8}F_{0.2}$) | *First step*: 900 °C, 45 h, 0 MPa (Open Ta-tube) ↓ *Second step*: 900 °C, 1 h, 1 GPa (sealed into a Ta-tube) |

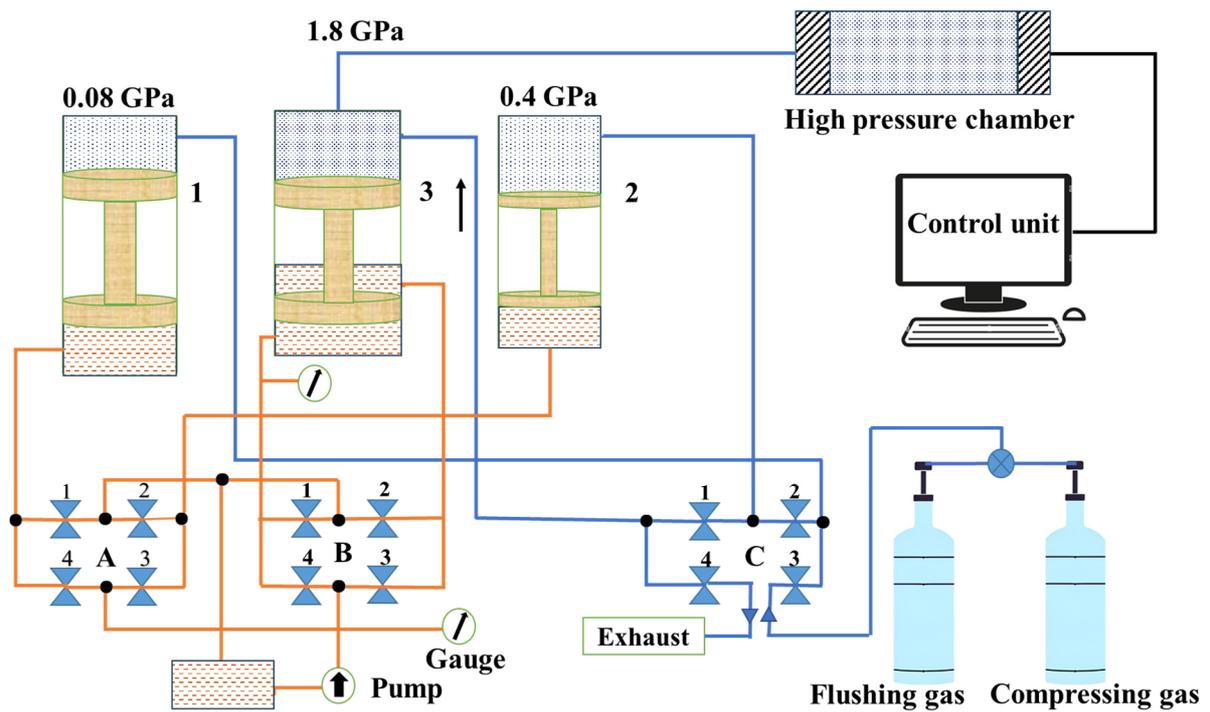

**Figure 1:** Block diagram of HP-HTS technique, which includes a three-stage oil-based compressor, a high-pressure chamber, and a control unit monitor. "A" and "B" depict a set of key valves to control the pressure through the oil pump for the three pistons: 1, 2 and 3 sustaining pressures of 0.08 GPa, 0.4 GPa and 1.8 GPa respectively. "C" represents a set of key valves to control the gas pressure for the three pistons [23].

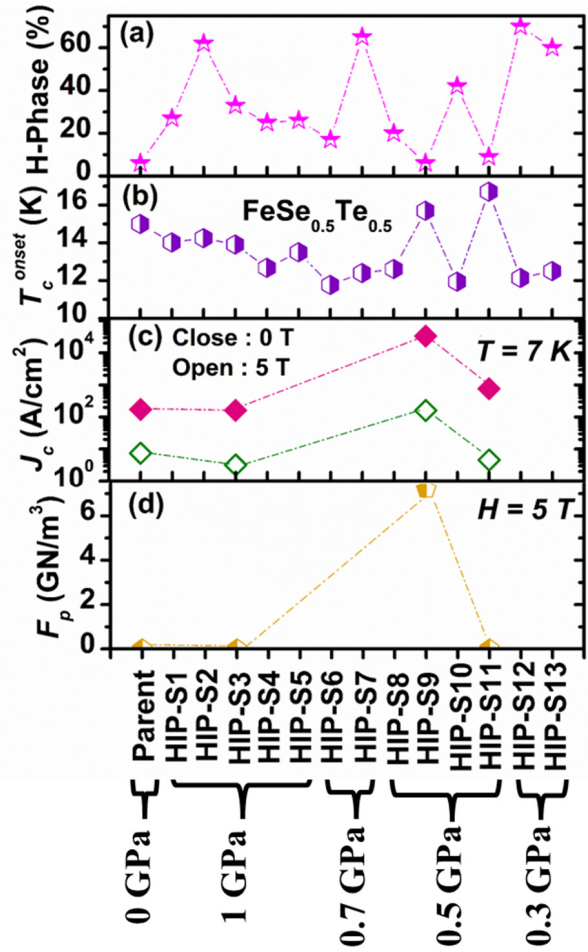

**Figure 2:** The synthesis pressure dependence of **(a)** hexagonal H-phase calculated from XRD patterns, **(b)** the $T_c^{onset}$, **(c)** the $J_c$ values at 0 T and 5 T, and **(d)** the calculated pinning force ($F_p$) at the applied magnetic field of 5 T [15] for various FeSe$_{0.5}$Te$_{0.5}$ prepared by HP-HTS.

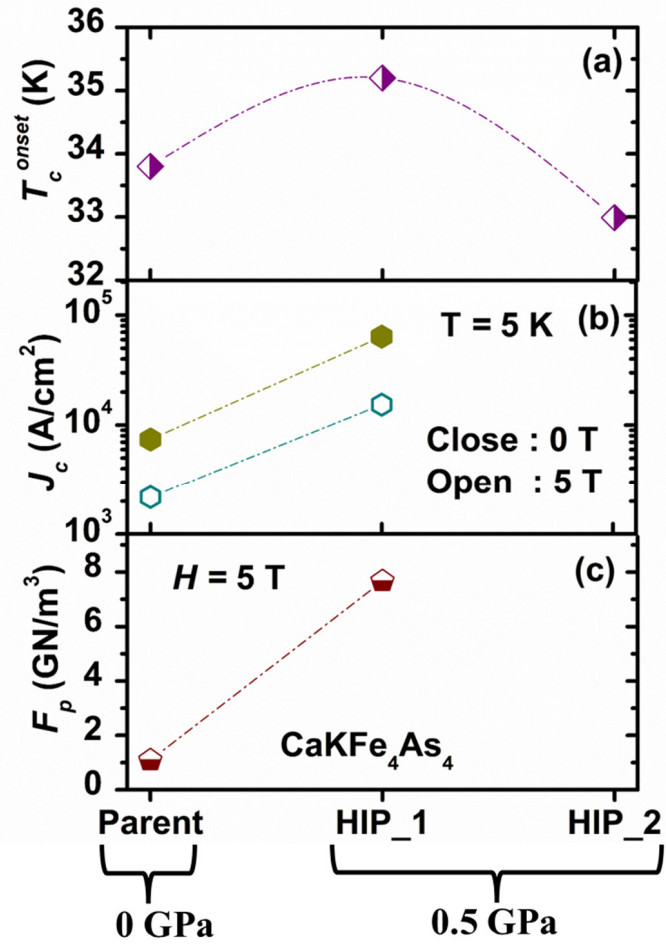

**Figure 3:** The variations of **(a)** the $T_c^{onset}$, **(b)** the $J_c$ value at 0 T and 5 T, **(c)** the calculated pinning force ($F_p$) at the applied magnetic field of 5 T [28],[29] for the prepared bulk CaKFe$_4$As$_4$ samples with respect to the synthesis pressure.

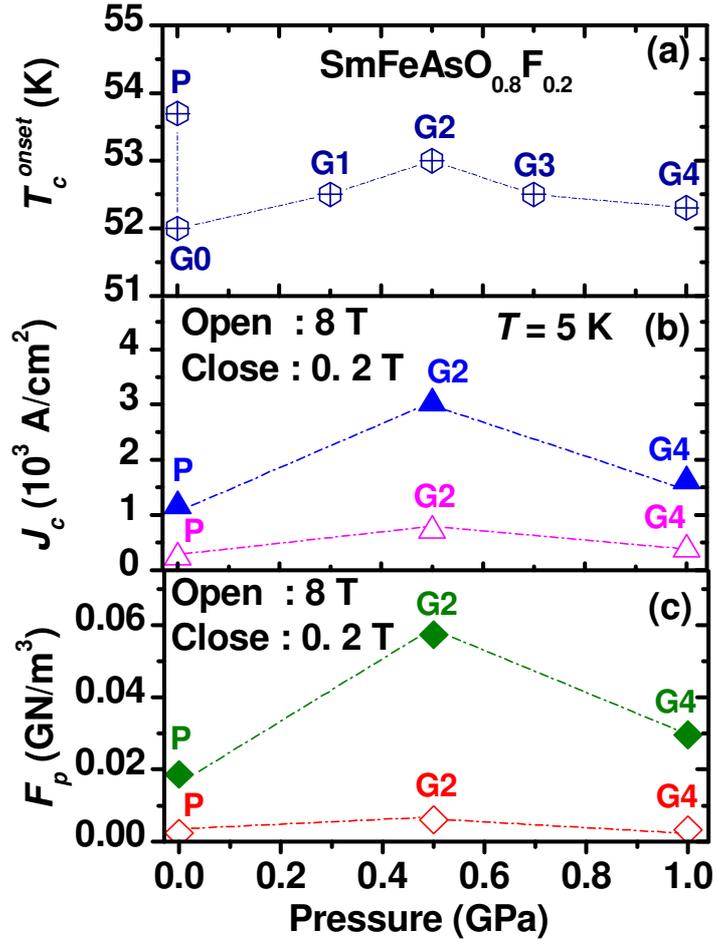

**Figure 4:** The synthesis pressure dependence of **(a)** the variations of the $T_c^{onset}$ for G0-G4 (G-batch) and parent (P) samples, and **(b)** the $J_c$ values at 0.2 T and 8 T, **(c)** the calculated pinning force ($F_p$) at 0.2 T, 8 T, and 5 K for G2, G4 and the parent (P) samples [24].